\documentclass[
twocolumn
 amsmath,amssymb,
 aps,
]{revtex4-2}

\usepackage{graphicx}% Include figure files
\usepackage{dcolumn}% Align table columns on decimal point
\usepackage{bm}% bold math
\usepackage[colorlinks=true, allcolors=blue]{hyperref}

\begin{document}

\preprint{APS/123-QED}

\title{Vibronics of multi-material nanopillared membranes and impact on the thermal conductivity}

\author{Lina Yang\textsuperscript{1} }
 %Lines break automatically or can be forced with \\
\author{Mahmoud I. Hussein\textsuperscript{2,3}}%
 \email{mih@colorado.edu}
\address{%
\textsuperscript{1}School of Aerospace Engineering, Beijing Institute of Technology, Beijing 100081, China
\\
\textsuperscript{2}Smead Department of Aerospace Engineering Sciences, University of Colorado Boulder, Boulder, Colorado 80303, USA
\\
\textsuperscript{3}Department of Physics, University of Colorado Boulder, Boulder, Colorado 80302, USA}%

\date{\today}% It is always \today, today,
             %  but any date may be explicitly specified

\begin{abstract} 
Atomic motion in nanopillars standing on the surface of a silicon membrane generates vibrons, which are wavenumber-independent phonons that act as local resonances. These vibrons couple with heat-carrying phonons traveling along the base membrane causing a reduction in the in-plane lattice thermal conductivity. In this work, we examine isolated silicon and gallium nitride nanopillars and for each compare the vibrons density of states (DOS) to those of phonons in an isolated uniform silicon membrane. We show that while the phonon-vibron DOS conformity across the full spectrum is a key factor in reducing the thermal conductivity of the assembled nanostructure, the presence of an intense vibron population at more dominant low frequencies plays a competing role. We report predictions from molecular dynamics simulations showing lower thermal conductivities for a silicon membrane with gallium-nitride nanopillars compared to a silicon membrane with silicon nanopillars.
\end{abstract}

                             % Classification Scheme.
%\keywords{Suggested keywords}%Use showkeys class option if keyword
                              %display desired
\maketitle

%\tableofcontents

\section{\label{sec:level1}Introduction:\protect\\  \lowercase{} }
The fast-advancing field of phononics targets the control of waves and vibrations in artificially structured materials~\cite{hussein2011preface,hussein2014preface,hussein2016preface,hussein2014dynamics}.~The subfield of nanophononics, also rapidly emerging, aims to apply phonon engineering concepts at the atomic scale~\cite{balandin2005nanophononics,li2012colloquium,volz2016nanophononics,hussein2020nanophononics,nomura2022review}.~Atomic-scale phonons are quantized vibrations, which are the main heat carriers in solids~\cite{levinshtein2001properties,kittel2005introduction}.~Nanostructuring is an effective way to manipulate nanoscale thermal transport because of the possibility of targeting characteristic length scales smaller than the mean free path (MFP) of phonons~\cite{esfarjani2011heat}. Because of its ability to reduce the lattice thermal conductivity in particular, nanostructures have been widely studied to improve the efficiency of thermoelectric (TE) energy conversion$-$a technology that has promising potential for industrial impact especially in the conversion of waste heat into electricity~\cite{chen2003recent,snyder2008complex,liu2012recent,biswas2012high}. The performance of TE materials is determined by a figure of merit $ZT = S^{2} \sigma  T/\kappa)$, where $S$ is Seebeck coefficient, $\sigma$ and $\kappa$ are the electrical conductivity and thermal conductivity, respectively, and $T$ is the average temperature. Some nanostructuring approaches aim to increase the value of $ZT$ by reducing the thermal conductivity while causing no or a minimum effect on the electronic properties~\cite{dresselhaus2007new,heremans2013thermoelectrics}. 

The process of nanostructuring is performed on a bulk crystal, rendering it a nanostructured material. Silicon (Si) has been viewed as an attractive candidate because it is a relatively inexpensive Earth abundant material that is nontoxic and has favorable mechanical properties.~However, in its bulk form it is a poor TE material because of its high lattice thermal conductivity ($\sim$150 W/m-K at 300 K)~\cite{weber1991transport}. Nevertheless, given silicon's high ratio of lattice thermal conductivity to electronic thermal conductivity, it is well suited for techniques that influence the lattice thermal conductivity. Examples include thin Si nanowires~\cite{yu1994electron} and membranes~\cite{bannov1995electron,balandin1998significant}, particularly when the free surfaces are rough~\cite{hochbaum2008enhanced,neogi2015tuning}. These reduced-dimension Si nanomaterials exhibit reduced group velocities in their phonon band structure (in their pristine form) or diffusive boundary scattering from the free edges (with surface roughness). More complex forms of nanostructuring have also been investigated, targeting phonon Bragg scattering from periodic nanoscale structural features such as repeated layers in a superlattice~\cite{barker1978study,yamamoto1994coherent,venkatasubramanian2000lattice,chen2005minimum,mcgaughey2006phonon,simkin2000minimum,ravichandran2014crossover} or an array of holes or inclusions in a nanophononic crystal~\cite{tang2010holey,yu2010reduction,davis2011thermal,he2011thermal,zen2014engineering,Minnich2014,yang2014extreme,yang2016thermoelectric,feng2016ultra,lee2017investigation,hu2018randomness,hussein2020nanophononics}.~A key limitation, however, is that nanoscale Bragg scattering requires almost defect-free conditions and near-perfect internal surfaces~\cite{lee2007lattice,Minnich2014,lee2017investigation}.

An alternate emerging strategy of nanophononics is based on the introduction of nanoresonators to a crystalline medium to generate intrinsic local resonances$-$or vibrons$-$in the phonon spectrum~\cite{davis2014nanophononic,wei2015phonon,honarvar2016spectral,honarvar2016thermal,xiong2016blocking,honarvar2018two,zhang2020novel,hussein2020nanophononics,wang2021anomalous,chen2022phonon,liu2023enhanced,li2023phonon,spann2023semiconductor}.~These vibrons couple with the phonons of the underlying host crystal to create group velocity reductions and mode localizations. These effects, which in principle may be tuned to cover the entire spectrum, cause a reduction in the in-plane thermal conductivity. Moreover, when the nanoresonators are in the form of nanopillars standing on a membrane, the electronic properties of the base membrane may remain intact$-$leading to an increase in the ZT value~\cite{spann2023semiconductor}. A nanostructure exhibiting atomic-scale local resonances is commonly referred to as a \textit{nanophononic metamaterial} (NPM)~\cite{davis2014nanophononic,Hussein2018Handbook,hussein2020nanophononics}.  

NPMs composed of a single type of material (with or without alloying) for both the base medium and the nannoresonator have been studied extensively, e.g., Refs.~\cite{davis2014nanophononic,wei2015phonon,honarvar2016spectral,honarvar2016thermal,xiong2016blocking,honarvar2018two,ma2018unexpected,zhang2020novel,liu2023enhanced}. For example, among the configurations investigated are suspended Si membranes with Si nanopillars standing on one or both surfaces~\cite{davis2014nanophononic,wei2015phonon,honarvar2016spectral,honarvar2016thermal,honarvar2018two,liu2023enhanced}, suspended Si membranes with Si nanowalls~\cite{honarvar2018two}, Si nanowires with silicon nanopillars extruded out of the core nanowire stem~\cite{xiong2016blocking}, and Si nanowires with a surface screw thread on the surface~\cite{zhang2020novel}. Other single-material NPMs considered are based on graphene, such as pillared graphene nanoribbons~\cite{ma2018unexpected}. \\
\indent Using different materials for the host medium (e.g., membrane) and the nanoresonator (e.g., nanopillar) in an NPM may create some challenges regarding the quality of the coupling at the interface between the materials, but may also offer some design flexibility for improvement of performance. Prior studies examined varying the atomic mass in the nanoresonator~\cite{ma2018unexpected,ma2021anomalous} or a different chemical material altogether compared to the material of the host medium~\cite{spann2023semiconductor,li2023phonon}. Spann et al. experimentally investigated a random arrangement of gallium nitride (GaN) nanopillars on a suspended Si membrane and also reported computational analysis of the same system but with a periodic arrangement of the nanopillars~\cite{spann2023semiconductor}. Their  results have shown that the material mismatch between the membrane and nanopillar does not prevent the resonance hybridizations to take place. \\ 
\indent In this work, we systemically investigate the thermal transport properties of nanopillared-membrane NPMs constructed from two types of materials (GaN nanopillars periodically arranged on a suspended Si membrane).~We select GaN for the nanoresonantor for numerous reasons. GaN is a key semiconductor due to featuring a wide electronic band gap, a quality highly desired in modern electronics~\cite{levinshtein2001properties,gaskill1996iii}. It has been extensively studied~\cite{wang1994diamond,Ponce1996}, also motivated also by its high thermal conductivity~\cite{sichel1977thermal} and good mechanical strength~\cite{levinshtein2001properties}.~Moreover, effective fabrication techniques of GaN nanostructures are well developed, e.g., see Refs.~\cite{bertness2006spontaneously,spann2017,spann2023semiconductor}.~Most important for NPMs is that~\textit{ab initio} calculations have previously predicted that most of the thermal conductivity of GaN is associated with phonons with MFPs greater than 200 nm~\cite{garg2018spectral}. In the present analysis, we show using equilibrium molecular dynamics (MD) simulations that NPMs with GaN nanopillars standing on a suspended Si membrane (GaN-on-Si) can cause larger reductions in the thermal conductivity than corresponding NPMs based on Si-on-Si. In addition to thermal conductivity prediction by MD simulations, we obtain the harmonic and anharmonic phonon band structures using lattice dynamics (LD) and spectral energy density (SED) calculations, respectively, to gain insights into the mechanisms of multi-material NPMs and the conditions needed for their ability to trigger stronger reducitons in the thermal conductivity compared to their single-material counterparts. 
%Phonon dispersion analyses by lattice dynamics shows that GaN pillars can introduce more local vibrational modes at lower frequencies. In addition, spectral energy density is calculated for GaN NPM, which show that the GaN pillars do alter phonon modes in Si membranes. The nonconformity factor of GaN pillar and Si membrane is compared with that of Si pillar and Si membrane to demonstrate the ability of GaN pillar on reducing thermal conductivity of Si membrane. Further, mode weight factor is computed for GaN NPM to illustrate the localization effect caused by pillars. Finally, the pillar height effect on reducing thermal conductivity of GaN NPM is investigated.

\section{Models and Methods}
\begin{figure*} [b!]
\includegraphics{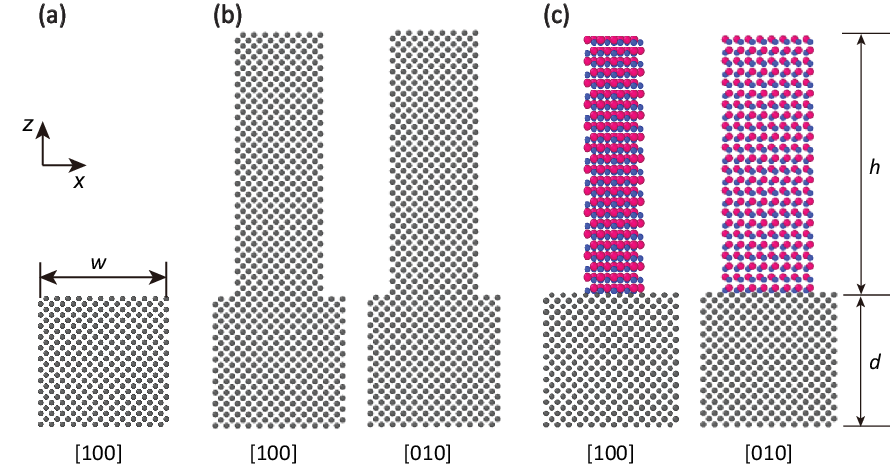}
\caption{\label{fig:fig1}~\textbf{Atomic models:} (a) Structure of Si membrane in [100] direction. (b) Si-on-Si NPM in [100] and [010] directions. (c) GaN-on-Si NPM in [100] and [010] directions. The Si membrane or base membrane is formed from $6\times 6 \times 6 $ CC. The Si nanopillar in (c) is constructed from $4\times 4 \times 12 $ CC. The GaN nanopillar in (b) is constructed from $4\times 4 \times 12 $ UC. The number of atoms in the GaN nanopillar is the same as that in Si nanopillar.}
\end{figure*}

The suspended Si membrane and Si nanopillars are created from the Si conventional cell (CC)  with lattice constant 0.5431 nm as a building block~\cite{davis2011thermal,davis2014nanophononic}. The GaN nanopillars are built from a unit cell (UC) of GaN with lattice constants $L_a$=0.319 nm, $L_b$=0.553 nm, $L_c$=0.52 nm in the $x$, $y$ and $z$ directions, respectively. The unit cell of GaN also contains 8 atoms which is the same as that in the Si CC. The simulation configurations of the three key models we consider, Si membrane, Si-on-si NPM, and GaN-on-Si NPM, are shown in Figs.~\ref{fig:fig1}(a),~\ref{fig:fig1}(b) and~\ref{fig:fig1}(c), respectively. The GaN is in wurtzite crystal structure. The width $W$ and thickness $t$ of the Si membrane unit cell, or the base membrane in an NPM unit cell, are both 3.26 nm (6 CC). The height $H$ of the nanopillar in the NPM unit cell is varied. Periodic and free boundary conditions are applied in the in-plane and out-of-plane directions, respectively. For comparison, the thermal conductivity of bulk Si and bulk GaN are also calculated. The computational cell of bulk Si and bulk GaN is set as $6\times 6 \times 6 $ CC and $13\times 8 \times 8 $ UC, respectively, to overcome any computational-domain size effects~\cite{schelling2002comparison,sellan2010size,li2010strain}.
The Tersoff potential is used to describe the interaction between Si atoms~\cite{tersoff1988empirical} and between the gallium (Ga) and nitrogen (N) atoms~\cite{nord2003modelling}.~The couplings between the Si and Ga atoms and between the Si and N atoms are obtained by following the combination rule~\cite{tersoff1989modeling}.~This combination approach has been applied to silicon carbide (SiC) systems, and the properties of SiC and its defects are well described~\cite{tersoff1989modeling}.~Equilibrium MD simulations~\cite{schelling2002comparison,yang2013reduction} are performed in LAMMPS~\cite{plimpton1995fast} at $T_{\text{R}}=300$ K.

Following the simulations, the thermal conductivity is calculated from the Green-Kubo formula~\cite{schelling2002comparison}
\begin{equation}
~~~~\kappa =\frac{1}{2Vk_{B} T^{2}} \int_0^{\infty} \langle \bm{J}(0)\cdot \bm{J}(\tau) \rangle \text{d} \tau,  \\~~~~ 
\label{eq:eq1}
\end{equation}
where $k_{B}$ is the Boltzmann constant, $V$ is the material system volume, and $T$ is the temperature. The time step in the simulations is set as 0.5 fs. First, a canonical ensemble MD with a Langevin heat reservoir is allowed to run for $6 \times 10^{5}$ steps (300 ps) to equilibrate the system at $T_{\text{R}}$. Then, a microcanonical ensemble (NVE) runs for  $8 \times 10^{6}$ steps (4 ns); meanwhile, the heat current is recorded at each step. At the end of the simulations, the thermal conductivity is calculated by Eq.~(\ref{eq:eq1}). The final results are averaged over twenty simulations with different initial conditions. The statistical error values are obtained following the method of Ref.~\cite{schelling2002comparison}.

%Note the open one in Eq.~(\ref{eq:one}).

To elucidate the underlying physics of the phonon transport in NPMs, the phonon dispersion of both NPMs and a corresponding uniform membrane are obtained by LD calculations using the GULP software pacakge~\cite{gale1997gulp}. To further examine the resonance hybridization mechanism as manifested in the anharmonic MD simulations, the SED~\cite{thomas2010predicting,larkin2014comparison,honarvar2016spectral} is also calculated directly from the MD simulation results by analysis of the velocities of the atoms as a function of space and time. As described in Ref.~\cite{larkin2014comparison}, the SED expression $\Phi'$ is given by
\begin{equation}
\Phi'(\bm{k},\omega)=\frac{1}{4\pi\tau_{0}}\sum_{\alpha=1}^{3}\sum_{b=1}^{n}\frac{m_b}{N}\vert \sum_{l=1}^{N}\int_{0}^{\tau_0}\dot{u}_{\alpha}\left(l,b,t\right) \text{e}^\Theta \text{d}t|^2,
\label{eq:eq2}
\end{equation}
where $\dot{u}_\alpha$ is the $\alpha$th component of the velocity
of the $b$th atom in the $l$th unit cell at time $t$, and $\Theta=i[\bm{k}\cdot\bm{r}_0\left(l,b\right)-\omega t]$. The parameters $m_b$ is the mass of the $b$th atom, $\tau_0$ is the  simulation time, and $\bm{r}_0$ is the equilibrium position vector
of the $l$th unit cell. Here, the unit of $\Phi'$ is J$\cdot$ps. The total number of unit cells is denoted by $N$ and the number of atoms in the unit cell is denoted by $n$.

\section{Results and Discussions}
\subsection{Thermal conductivity}
Our calculated thermal conductivity for bulk Si at 300 K is 238.3$\pm$3.7 W/m-K, which does not deviate significantly from similar predictions in the literature~\cite{lee2007lattice}. The thermal conductivity of bulk GaN is calculated as 180.83$\pm$0.73 W/m-K, which is very close to the MD prediction of 185 W/m-K reported in Ref.~\cite{zhou2009towards}.  The thermal conductivity of bulk Si and GaN by  equilibrium MD simulations deviate from experimental results due to the inaccuracies of the semi-empirical potentials used in the simulation models and the impurity of examined samples in experiments. Nevertheless, these discrepancies do not affect our comparison of the results for different nanostructured NPMs among themselves and with a corresponding uniform Si membrane since the same simulation method and potential parameters are used in all cases.  

\begin{figure*} [b]
\includegraphics{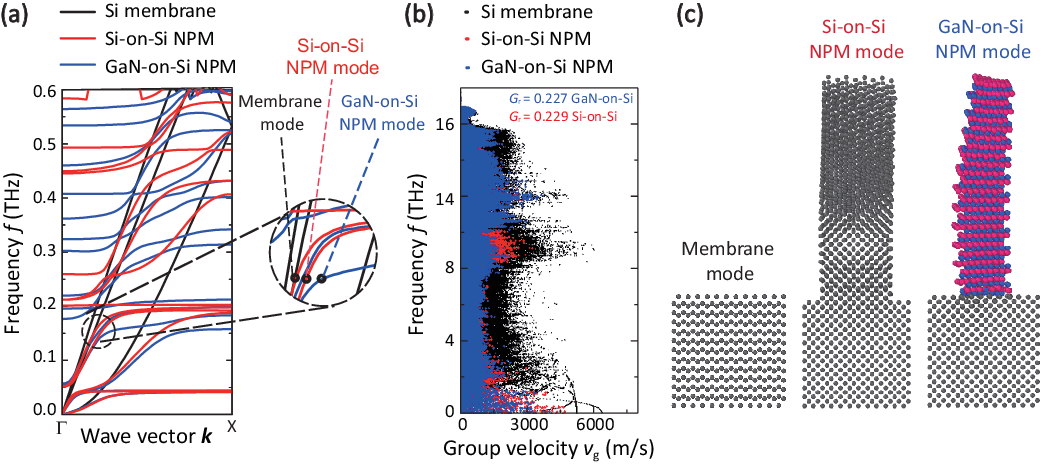}% Here is how to import EPS art
\caption{\label{fig:fig2}~\textbf{Resonant hybridization phenomenon:} (a) Phonon dispersion of Si membrane (black line), Si-on-Si NPM (red line) and GaN-on-Si NPM (blue line) shown in Fig.~\ref{fig:fig1}. The zoom-in is on the hybridization region circled by black dash line in (a); the solid black circles label a phonon mode for the Si membrane and hybridized modes in the Si-on-Si NPM and GaN-on-Si NPM, respectively. (b) Group velocities of phonon modes in Si membrane (black dots), Si-on-Si NPM (red dots), and GaN-on-Si NPM (blue dots). The average group velocity ratio $G_r$ for the Si-on-Si NPM and GaN-on-Si NPM is 0.229 and 0.227, respectively, for modes with frequency less than 6 THz. (c) Mode-shape images of the atomic displacement for the highlighted phonon mode in the Si membrane and hybridized modes in the Si-on-Si NPM and GaN-on-Si NPM, respectively. }
\end{figure*}

\begin{figure*}
\includegraphics{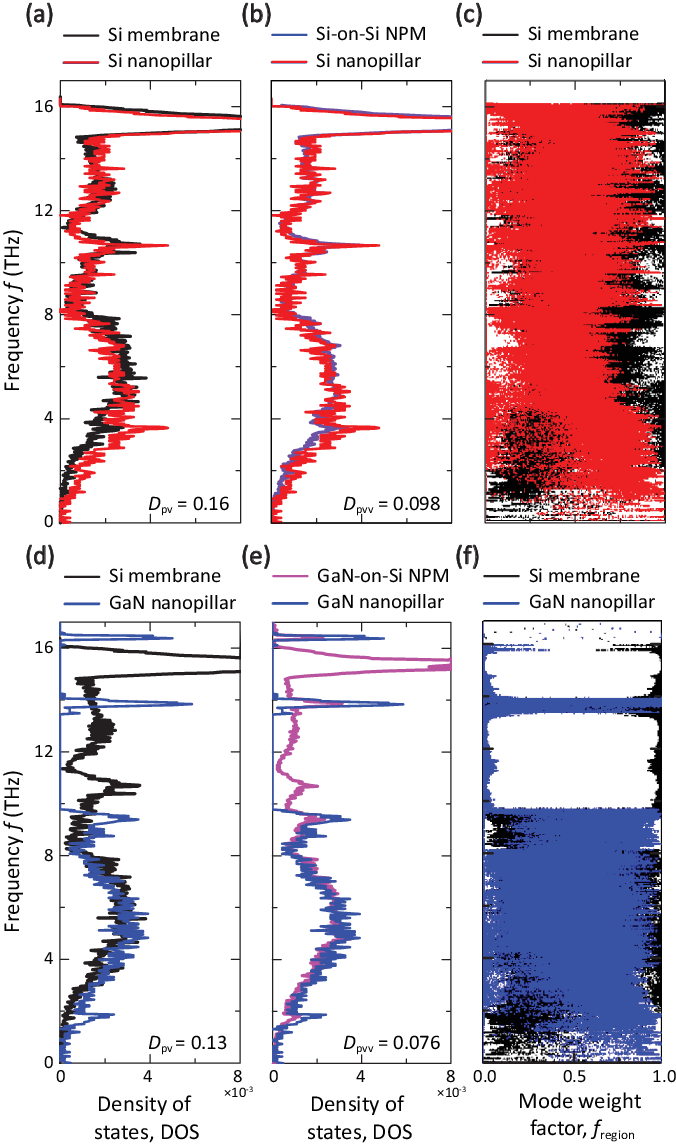}% Here is how to import EPS art
\caption{\label{fig:fig3}~\textbf{Density of states and mode weight factor:} (a) DOS of GaN nanopillar vibrational modes (blue line) and Si membrane phonon modes (black line). (b) DOS of GaN nanopillar vibrational modes (blue line) and GaN NPM modes (pink line). (c) Mode weight factor of GaN nanopillar (blue dots) and Si membrane (black dots) for GaN NPM. For this combination, the mode weight factor ratio is $f_r$ is 0.95 for the modes with frequency less than 10 THz. (d) DOS of Si nanopillar vibrational modes (red line) and Si membrane phonon modes (black line). (e) DOS of Si nanopillar vibrational modes (red line) and Si NPM modes (purple line). (f) Mode weight factor of Si nanopillar (red dots) and base Si membrane (black dots) for Si NPM. For this combination, the mode weight factor ratio is $f_r$ is 0.98 for the modes with frequency less than 10 THz. The structure of the Si membrane, Si NPM, and GaN NPM are shown in Fig.~\ref{fig:fig1}. The nonconformity factors $R_{pv}$ provided in (a),(b),(d) and (e) are calculated using DOS with frequency less than 8 THz. } 
\end{figure*} 

\begin{figure*} [b]
\includegraphics{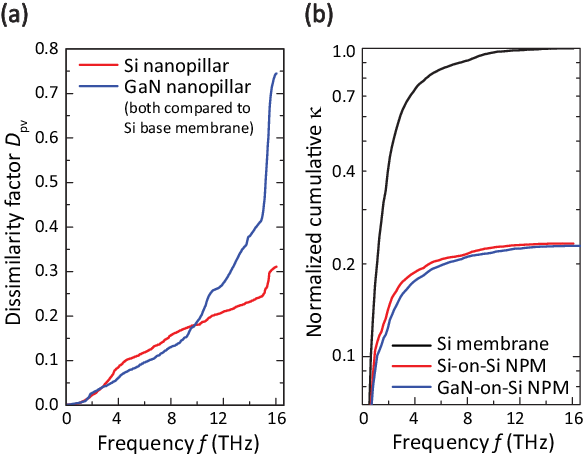}% Here is how to import EPS art
\caption{\label{fig:fig4}~\textbf{DOS dissimilarity analysis:} (a) Dissimilarity factor for Si membrane and GaN nanopillar (blue solid line), Si membrane and Si nanopillar (red solid line). The dissimilarity factor $D_{pv}$ is calculated based on the DOS in Figs.~\ref{fig:fig3}(a) and~\ref{fig:fig3}(d). (b) Cumulative thermal conductivity (normalized with respect to the maximum) value versus frequency for Si-on-Si NPM (red line) and GaN-on-Si NPM (blue line) calculated by the C-H model.}
\end{figure*}

\begin{figure*} [b]
\includegraphics{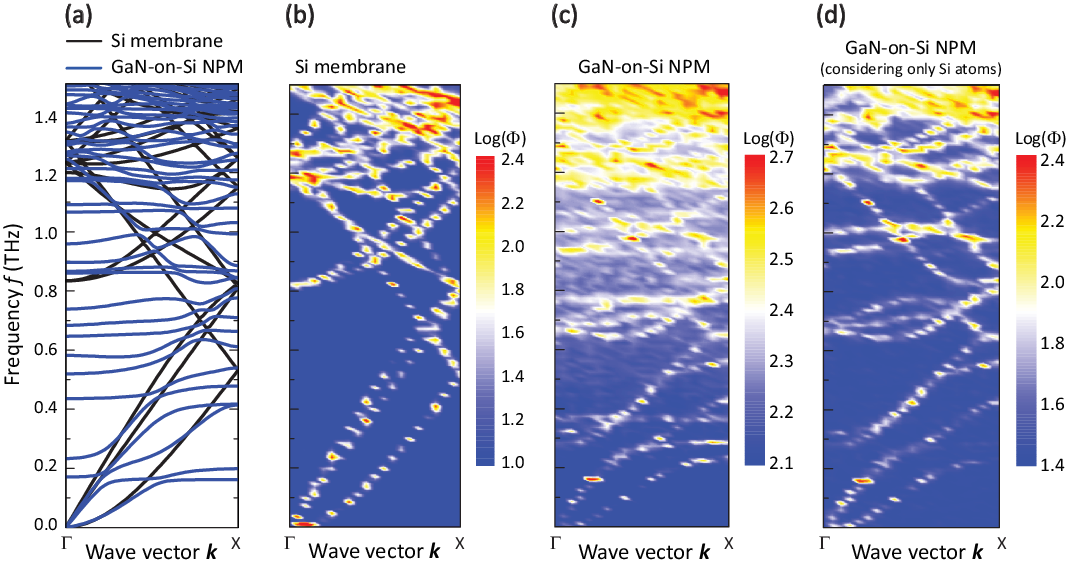}% Here is how to import EPS art
\caption{\label{fig:fig5}~\textbf{Harmonic and anharmonic phonon band structure:} (a) Phonon dispersion of Si membrane and GaN-on-Si NPM by LD. SED spectrum of (b) Si membrane and (c) GaN-on-Si NPM. (d) SED spectrum of GaN-on-Si NPM by considering only the atoms in the Si base membrane portion. In the GaN-on-Si NPM, the GaN nanopillar with a height of 1.56 nm is constructed from 4$\times$4$\times$3 UC, and the base Si membrane is the same as that in Fig.~\ref{fig:fig1}.}
\end{figure*}

\begin{figure*}[h!]
\includegraphics{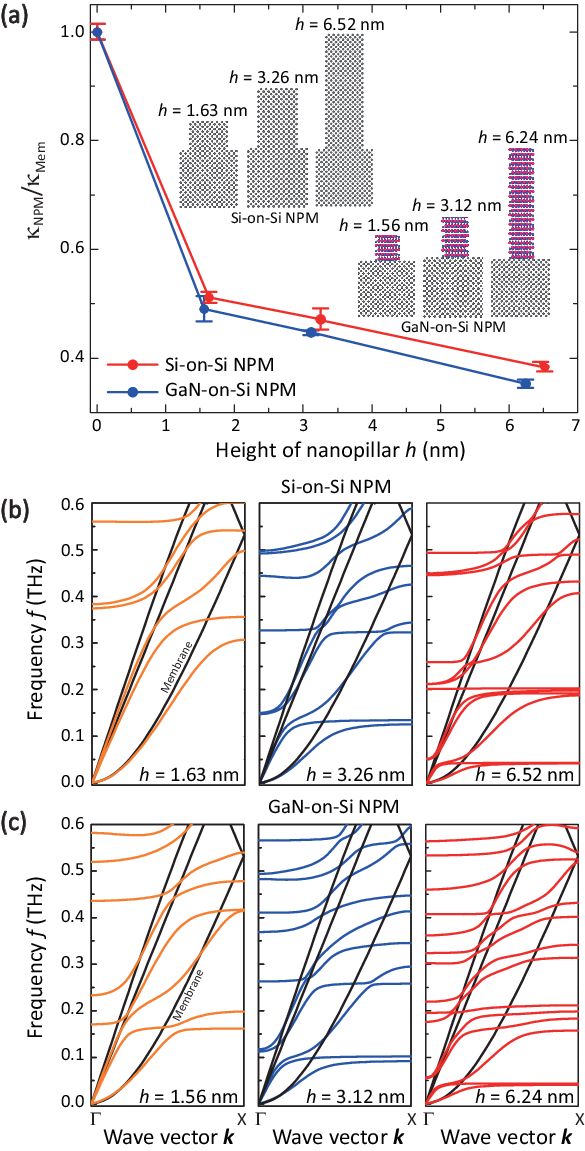}% Here is how to import EPS art
\caption{\label{fig:fig6}~\textbf{Effect of nanopillar height:} (a) Ratio of the thermal conductivity of GaN-on-Si NPM ($\kappa_{\text{NPM}}$) to that of Si membrane ($\kappa_{\text{Mem}}$) (blue dots), and ratio of thermal conductivity of Si-on-Si NPM ($\kappa_{\text{NPM}}$) to that of Si membrane ($\kappa_{\text{Mem}}$) (red dots). The height of GaN nanopillars is set as 3 UC (1.56 nm), 6 UC (3.12 nm) and 12 UC (6.24 nm). The height of Si nanopillars is set as 3 UC (1.63 nm), 6 UC (3.26 nm) and 12 UC (6.52 nm). The thickness of Si membrane is set as 6 CC (3.26 nm). The nanostructures of GaN-on-Si NPM and Si-on-Si NPM are also shown. (b) Phonon dispersion of GaN-on-Si NPM with different nanopillar heights in (a). (c) Phonon dispersion of Si-on-Si NPM with different nanopillar heights in (a).}
\end{figure*}

The thermal conductivity predictions for the Si membrane, Si-on-Si NPM, and GaN-on-Si NPM models shown in Fig.~\ref{fig:fig1} are listed in Table \ref{table:table1}, and $\kappa$ as a function of time is shown in Fig.~\ref{fig:figA1}. Here, the Si and GaN nanopillar sizes are chosen to be $4\times 4 \times 12 $ CC and $4\times 4 \times 12 $ UC, respectively. The last row of Table~\ref{table:table1} shows the ratio of the thermal conductivity of the Si-on-Si NPM and GaN-on-Si NPM to that of Si membrane. The thermal conductivity of Si-on-Si NPM and GaN-on-Si NPM is 38\% and 35\% of that of the Si membrane, respectively. The reduction of $\kappa$ of the Si-on-Si NPM is relatively consistent with predictions reported in previous studies~\cite{honarvar2016thermal,honarvar2018two,hussein2020nanophononics}. Importantly, we observe that the GaN nanopillar has a stronger effect on the reduction of the thermal conductivity than the Si nanopillar.~This is despite of the fact that these are two different materials with dissimilar density of states (DOS)~\cite{honarvar2018two}, not to mention the presence of a dissimilar matertial interface between the GaN nanopillar and the Si base membrane.

\begin{table}[t]
\centering
\caption{Thermal conductivity of Si membrane, Si NPM, and GaN NPM in Fig.~\ref{fig:fig1}. The values are normalized with respect to the thermal conductivity of a corresponding Si membrane $\kappa_{\text{Mem}}$.} % title of Table
%\centering % used for centering table
\begin{tabular}{m{6cm}| m{6cm}| m{6cm}| m{6cm}} % centered columns (4 columns)
\hline %inserts double horizontal lines
\multicolumn{1}{c|}{Structure}    &  \multicolumn{1}{c|}{Si membrane}    &  \multicolumn{1}{c|}{Si-on-Si NPM}  &  \multicolumn{1}{c}{GaN-on-Si NPM}  \\  % inserts table
%heading
\hline % inserts single horizontal line
\multicolumn{1}{c|} {$\kappa$ (W/m-K)} & \multicolumn{1}{c|} {40.6$\pm$0.6} & \multicolumn{1}{c|} {15.6$\pm$0.4} & \multicolumn{1}{c} {14.3$\pm$0.3} \\ % inserting body of the table
\hline
\multicolumn{1}{c|} {$\kappa$/$\kappa_{\text{Mem}}$}& \multicolumn{1}{c|} {1.0} & \multicolumn{1}{c|}  {0.38} & \multicolumn{1}{c}  {0.35} \\
\hline %inserts single line
\end{tabular}
\label{table:table1} % is used to refer this table in the text
\end{table}

\subsection{Harmonic phonon dispersion}
To gain insights into the underlying mechanism behind the reduction of the thermal conductivity and why the reduction is higher with a GaN nanopillar rather than a Si nanopillar, we examine the phonon band structures. The three systems shown in Fig.~\ref{fig:fig1} are again examined, namely the Si membrane, Si-on-Si NPM, and GaN-on-Si NPM. First, we calculate the phonon dispersion following the harmonic approximation~\cite{gale1997gulp}; see Fig.~\ref{fig:fig2}(a). It is observed that the nanopillars introduce vibron modes that couple and hybridize with the phonon modes in the underlying Si base membrane, causing the latter modes to flatten which in turn manifests as a reduction of the group velocities. This effect along with the energy localization of the modes within the nanopillar portion of the nanostructure causes a reduction in the in-plane lattice thermal conductivity~\cite{davis2014nanophononic,wei2015phonon,honarvar2018two}. A critical observation is that there are more vibrons at low frequencies in the GaN-on-Si NPM than that in Si-on-Si NPM, which is significant given the usually more dominant role of low-frequency phonons in thermal transport in general~\cite{yang2017thermal}. 

The full-spectrum phonon group velocities of the Si membrane, Si-on-Si NPM, and GaN-on-Si NPM are plotted in Fig.~\ref{fig:fig2}(b) for all the wave vectors along the $\Gamma$-X direction. The phonon group velocities of the two NPMs are clearly reduced compared with that of the uniform Si membrane. Moreover, the GaN nanopillar is shown to have a stronger effect on these group velocity reductions at low frequencies, compared to the Si nanopillar. To quantify the reduction of the group velocities across a particular frequency range, we calculate the average group velocity, defined as
\begin{equation}
G=[1/(n_k n_m)]\sum_{k}^{n_k}\sum_{m}^{n_m}v_g(k,m),
\label{eq:eq3}
\end{equation}
where $v_g(k,m)$ is the group velocity, $n_k$ is the total number of wavenumbers considered along the $\Gamma$-X direction, and $n_m$ is the total number of dispersion branches within the frequency range of interest. We then calculate the ratio $G_r=G_{\text{NPM}}/G_{\text{Mem}}$, where $G_{\text{NPM}}$ and $G_{\text{Mem}}$ denote the average group velocity for NPM and membrane, respectively. The value of $G_r$ for the Si-on-Si NPM and the GaN-on-Si NPM are 0.229 and 0.227, respectively, specifically for modes with frequency less than 6 THz. This indicates that the average group velocity in this low frequency range is more strongly reduced in the GaN-on-Si NPM than that in Si-on-Si NPM. 

To show the difference between a standard phonon mode in a Si membrane and hybridized modes in a Si-on-Si NPM and a GaN-on-Si NPM, the three modes labeled by black circles in the zoom-in in Fig.~\ref{fig:fig2}(a) are investigated. The mode-shape images of these three modes, which show the atomic displacements calculated by adding the corresponding component of its eigen vector to its equilibrium position with a 200 magnification for visualization, are shown in Fig.~\ref{fig:fig2}(c). It is clearly seen that for the hybridized modes in the two NPMs, the atomic motion in the base membrane potion is suppressed, while the motion of atoms in the nanopillar portion is significant. This confirms that concentration of modal localization in the nanopillars at the expense of motion in the base membrane is possible at the vicinity of phonon-vibron hybridization zones. This behavior is consistent with the unit-cell level reduced group velocities. 

Our next analysis examines the DOS since it was shown in the literature that the degree of conformity between the isolated base membranes and nanopillar DOS distribution is an important factor for maximising the impact of the local resonators~\cite{honarvar2018two,hussein2020nanophononics}. Figures.~\ref{fig:fig3}(a) and~\ref{fig:fig3}(b) compares the DOS of the Si nanopillar vibron modes with that of the phonon modes of the Si base membrane and the overall Si-on-Si NPM phonon modes~\footnote{Whenever the DOS is reported for a full NPM unit cell, it is calculated based on all modes in the Brillouin zone}, respectively. We observe that for the latter comparison, there is a very good match along the entire frequency spectrum. Similarly, Figs.~\ref{fig:fig3}(d) and~\ref{fig:fig3}(e) compares the DOS of the GaN nanopillar vibron modes with that of the Si base membrane and the overall GaN-on-Si NPM phonon modes, respectively. Here, however, for the latter comparison we find a matching up to 8 THz only. Furthermore, in Fig.~\ref{fig:fig3}(d) within the range $8\le f\le12$ THz, we observe a drop in the peak DOS within the isolated GaN nanopillar spectrum compared to the isolated base membrane spectrum. This drop is not seen in Fig. Fig.~\ref{fig:fig3}(a), since both the nanopillar and the base membrane are composed of the same material$-$thus expected to have a highly confirming DOS distribution across the full spectrum. The lower $\kappa$ of the GaN-on-Si NPM compared to the Si-on-Si NPM is therefore attributed to this downward shifting of the DOS of the GaN nanopillar compared to the Si nanopillar, considering that in general low frequency phonons contribute more to the $\kappa$. 

To further quantify the unique vibron-phonon DOS correlations for the Ga-on-Si NPM compared to the Si-on-Si NPM, we compute a \textit{dissimilarity factor} $D_{\text{pv}}$ defined as
\begin{equation}
D_{\text{xv}}(\omega)=\sum_{\omega_i \leq \omega} \vert DOS_{\text{x}}(\omega_i)-DOS_{\text{v}}(\omega_i) \vert,
\label{eq:eq4}
\end{equation}
where $DOS_{\text{v}}$ is the DOS of the isolated nanopillar vibrons ($\text{v}$), and $DOS_{\text{x}}$ is the DOS of the isolated base membrane phonons ($\text{x}=\text{p}$) or the DOS of the full NPM which includes the coupling of the phonons and vibrons ($\text{x}=\text{pv}$). The higher the dissimilarity factor at a given frequency, the more dissimilar are the DOS of the two entities being compared. The computed $D_{\text{xv}}$ values are shown in Fig.~\ref{fig:fig4}(a). Surprisingly, We observe that the DOS of the GaN nanopillar is less dissimilar with that of Si base membrane within the 4 to 10 THz range compared to the Si nanopillar, implying more intense hybridizations for the GaN-on-Si case. The value of $D_{\text{xv}}$($\omega=8$THz) are listed in Figs.~\ref{fig:fig3}(a),~\ref{fig:fig3}(b),~\ref{fig:fig3}(d) and~\ref{fig:fig3}(e) for each case. In addition, the $\kappa$ of the Si membrane, Si-on-Si NPM, and GaN-on-Si NPM are also calculated by Callaway-Holland (C-H) model using parameters in Ref.~\citep{davis2014nanophononic}. The accumulative $\kappa$ normalized by the $\kappa$ of the uniform Si membrane versus frequency are shown in Fig.~\ref{fig:fig4}(b). The thermal conductivity obtained by the C-H model for the GaN-on-Si NPM is smaller than that of the Si-on-Si NPM, which is consistent with the equilibrium MD/Green-Kubo predictions reported above.  

The eigenvector mode localization within the unit cell is investigated further using the notion of a \textit{mode weight factor} $f_{\text{Region}}$ to quantify “regional" localization for mode ($\bm{k},\omega$)~\cite{honarvar2018two}. This quantity is defined as 
\begin{equation}
f_{\text{Region}}(\bm{k},\omega)=\sum_{i=1}^{N_{\text{Region}}}\sum_{j=1}^{3}\phi_{ij}^{*}(\bm{k},m)\phi_{ij}(\bm{k},m),
\label{eq:eq5}
\end{equation}
where $\phi_{ij}(\bm{k},m)$ is the normal eigenvector component corresponding to atom $i$ and direction $j$ and $N_{\text{Region}}$ represents the number of atoms in a spatial ``region" in the unit cell. The mode weight factor represents the relative contribution of each spatial region of the system (base membrane or nanopillar) to the overall mode shape~\cite{bodapati2006vibrations,hu2012si,honarvar2018two}. The factor $f_{\text{Region}}$ ranges from 0 to 1. When the $N_{\text{Region}}$ is set to be the total number of atoms of the full unit cell, then $f_{\text{Region}}=$1. The computed quantities $f_{\text{Pillar,Si}}$ and $f_{\text{Mem}}$ for the Si-on-Si NPM are shown in Fig.~\ref{fig:fig3}(c) and $f_{\text{Pillar,GaN}}$ and $f_{\text{Mem}}$ for the GaN-on-Si NPM are shown in Fig.~\ref{fig:fig3}(f). The quantity $f_{\text{Mem}}$ corresponds to the base membrane region of the NPM. These results show that the GaN nanopillar has a larger mode weight factor than the Si nanopillar for modes in the frequency range $4\le\omega\le8$ THz. Furthermore, we define the mode weight factor ratio $f_r$ as 
\begin{equation}
f_r=\sum_{\bm{k}}\sum_{\omega}f_{\text{Mem}}(\bm{k},\omega)/\sum_{\bm{k}}\sum_{\omega}f_{\text{Pillar}}(\bm{k},\omega).
\label{eq:eq6}
\end{equation}
The lower the $f_r$ value, the higher the degree of eigenvector mode localization in the nanopillar. The $f_r$ of the Si-on-Si NPM and GaN-on-Si NPM as shown in Figs.~\ref{fig:fig3}(c) and~\ref{fig:fig3}(f) are 0.98 and 0.95, respectively, for all modes with frequency less than 10 THz. This indicates that the GaN nanopillar experiences a larger degree of eigenvector mode localization in this lower frequency range compared to the Si nanopillar, consistent with the superior thermal conductivity reduction of the GaN-on-Si NPM compared to the Si-on-Si NPM.

\subsection{Anharmonic phonon dispersion}
To demonstrate that the resonant hybridization mechanism exists and unfolds within the MD simulations of the current models that we run for thermal conductivity prediction, the SED phonon spectrum is computed directly from a separate set of MD simulations~\cite{larkin2014comparison,thomas2010predicting,honarvar2016spectral}. In these latter simulations, we consider 40 NPM unit cells in the $x$ direction and only 1 NPM unit cell in the $y$ and $z$ directions, and we enforce NVE conditions over 2$\times10^6$ time steps at 300 K. The results for the Si membrane and a GaN-on-Si NPM are given in Figs.~\ref{fig:fig5}(b) and~\ref{fig:fig5}(c), respectively. For this GaN-on-Si NPM, the GaN nanopillar with a height of 1.56 nm is constructed by 4$\times$4$\times$3 UC, and the base Si membrane is the same as that in Fig.~\ref{fig:fig1}. For comparison, the corresponding harmonic phonon dispersion curves calculated by LD are shown in Fig.~\ref{fig:fig5}(a). Because phonon modes with higher frequency are difficult to be distinguished in SED spectrum, here only the low frequency phonon modes in the range $0\leq f \leq 1.5$ THz are shown. In the SED spectrum of the GaN-on-Si NPM, the flat bands characteristic of resonant hybridization is clearly visible beginning from approximately 0.2 THz and upwards. To highlight the extent at which the GaN nanopillar vibrons affect the heat-carrying phonon modes in the Si base membrane, the SED spectrum of the GaN-on-Si NPM considering only the atoms in the Si base membrane portion is calculated and shown in Fig.~\ref{fig:fig5}(d). Comparing the SED spectrum in Figs.~\ref{fig:fig5}(b) and~\ref{fig:fig5}(d), we find that the GaN nanopillar vibrons significantly alter the fundamental nature of the phonons in the Si base membrane by appearing as flat bands that couple withe original phonon dispersion causing distinct avoided crossings throughout the spectrum.

\subsection{Effect of nanopillar height}
Finally, the effect of the nanopillar height on the thermal conductivity of the NPM is investigated for both the Si-on-Si and GaN-on-Si configurations. It has been shown that the larger the volume ratio of the nanopillar to the base membrane, the larger the reduction in $\kappa$ because of increased vibron-to-phonon ratio~\cite{honarvar2018two}. This implies that the larger the nanopillar height, the lower the thermal conductivity. Here we examine the effect of nanopillar material choice on this behavior. The thickness of Si base membrane is set as 6 CC (3.26 nm), while the height of the GaN nanopillar is set as 3 UC (1.56 nm), 6 UC (3.12 nm) and 12 UC (6.24 nm), respectively.
The ratio of the thermal conductivity of the GaN-on-Si NPM ($\kappa_{\text{NPM}}$) to that of the Si membrane ($\kappa_{\text{Mem}}$ is shown in Fig.~\ref{fig:fig6}(a). For comparison, the ratio of the thermal conductivity of Si-on-Si NPM to that of the Si membrane is also plotted (the effect of the nanopillar height on this system has been investigated extensively in Ref.~\cite{honarvar2018two}). The modeled structure of each NPM is shown as insets. The results show a stronger decrease in $\kappa$ for the GaN-on-Si NPM compared to the Si-on-Si NPM, with the difference in the level of reduction increasing as the nanopillar height is increased. For further insight, the harmonic phonon band structure of each NPM for the various nanopillar heights are shown in Figs.~\ref{fig:fig6}(b) and~\ref{fig:fig6}(c). It is observed that the local resonances start a lower frquencies for the GaN-on-Si NPM compared to the Si-on-Si NPM, with a stronger drop in the vibron frequencies with nanopillar height for the GaN-on-Si system. In conclusion, an increase in nanopillar height is more effective in reducing the thermal conductivity when the nanopillar is made out of GaN rather than Si.

\section{Conclusion}
We systemically studied the phonon properties and the thermal conductivity of two types of NPMs, namely a suspended Si membrane with Si nanopillars and a suspended Si membrane with GaN nanopilllars. In previous research, it was shown that it is beneficial to ensure a maximum frequency-dependent conformity between the nanoresonator vibron DOS and the base-membrane phonon DOS to maximize the resonant hybridization effect across the full spectrum~\cite{honarvar2018two}. This criterion is valid for a single-material NPM and it was shown that the level of DOS conformity may be improved by both optimal sizing and shaping of the nanoresonator. In our current investigation, we demonstrated that by having a nanoresonator material that is more dense that the base membrane material (e.g., GaN compared to Si), we can in fact further improve the impact of the nanoresonators on the thermal conductivity reduction. This is because the more vibrons being populated in the lower end of the frequency spectrum  where most of the heat is carried, the stronger the thermal conductivity reduction. To add further insight into this aspect, our investigation was complemented with a series of both harmonic and anharmonic phonon property calculations including the calculation of a dissimilarity factor (Fig. \ref{fig:fig4}) that quantifies the level of frequency-dependent difference between the DOS of the vibrons and the DOS of the phonons. It was shown that the dissimilarity factor for the GaN-on-Si NPM is lower than that of Si-on-Si at frequencies lower than 10 THz. Lastly, the impact of the height of the nanopillar in the NPM unit cell was explored indicting a favorable trend with nanopillar height for the GaN-on-Si NPM compared to the Si-on-Si NPM.

\begin{acknowledgments}
This research was partially supported by the National
Science Foundation (NSF) CAREER Grant No. 1254931.~This work utilized the Janus supercomputer, which was supported by the National Science Foundation (USA) (Award No. CNS-0821794) and the University of
Colorado Boulder. The Janus supercomputer was a joint effort
of the University of Colorado Boulder, the University of Colorado Denver, and the National Center for Atmospheric
Research (USA).
\end{acknowledgments}

\appendix*
%\section*{APPENDIX}

\renewcommand{\thefigure}{A\arabic{figure}}
\renewcommand{\thetable}{A\arabic{table}}
\renewcommand{\theequation}{A\arabic{equation}}

\setcounter{figure}{0}    
\setcounter{table}{0}    
\setcounter{equation}{0}

%\appendix

\section{Thermal conductivity prediction by the Green-Kubo method}
\label{CellVolF}
\begin{figure}[h!]
\includegraphics{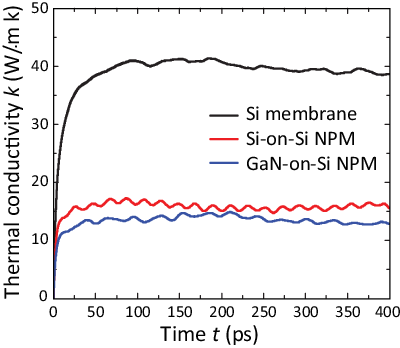}% Here is how to import EPS art
\caption{\label{fig:figA1}~\textbf{Thermal conductivity prediction:} (a) Thermal conductivity versus time at 300 K. The black, red, and blue curves correspond to the thermal conductivity of Si membrane, Si-on-Si NPM, and GaN-on-Si NPM in Fig.~\ref{fig:fig1}, respectively. The thermal conductivity of the GaN-on-Si NPM is lower than that of the Si-on-Si NPM.  }
\end{figure}

The time-dependent function of the thermal conductivity as obtained by equilibrium MD simulations followed by the Green-Kubo method is shown in Fig.~\ref{fig:figA1} for the three key models considered in this investigation. The results show excellent convergence and provide a clear demonstration of a reduction in the thermal conductivity by the addition of a nanopillar over a suspended Si membrane in the computational unit cell, with a stronger reduction for the GaN nanopillar compared to the Si nanopillar.

% The \nocite command causes all entries in a bibliography to be printed out
% whether or not they are actually referenced in the text. This is appropriate
% for the sample file to show the different styles of references, but authors
% most likely will not want to use it.
%\nocite{*}

\bibliography{Si-membrane_GaN-pillar}% Produces the bibliography via BibTeX.

\end{document}